\documentclass[12pt]{article}
\begin{document}
\noindent\textsc{Z. Haba},
\textit{Feynman Integral and Random Dynamics in Quantum Physics}.
Dordrecht: Kluwer Academic, 1999. 364 p., price 135 Euro (hc)
(Mathematics and Its Applications; 480). ISBN 0-7923-5735-3.

Useful as the Feynman or path integral may be, it seldom is a real integral to begin with, formally expressing a particle's propagator as a sum of amplitudes $e^{iS/\hbar}$ (with $S$ the classical action) over all continuous paths from initial to endpoint. Thus, the `integrand' has unit modulus everywhere in the infinite-dimensional path space, and physically irrelevant trajectories have their contributions suppressed only by destructive interference, i.e., by rapid oscillations of the amplitude. This is unsatisfactory not only for the mathematician: it is also an obstacle to (numerical) approximation of the path integral, and thus the reason why many problems in quantum physics in fact still are too hard for us.

The best-known attempt at a resolution formulates the Schr\"odinger equation in imaginary time, where it becomes a diffusion equation amenable to Wiener integration. This route is particularly attractive, since the quantum-statistical operator $e^{-H/T}$ (with $H$ the Hamiltonian and $T$ the temperature in energy units) is---incidentally or profoundly---an imaginary-time evolution operator. Many static problems can thus be analyzed, but a continuation back to real time is usually only possible for exact or perturbative analytic results (i.e., still better than claimed in the author's foreword, stating that imaginary-time methods never lead to dynamic information).

Positive amplitudes, and hence a probabilistic interpretation, can also be obtained by keeping the time real, but rotating the \emph{space} variables over $\sqrt{i}=e^{i\pi/4}$. This is the approach chosen in Haba's book. Preliminary chapters deal with quantum and classical mechanics, probability theory and stochastic calculus, and the conventional approach to path integration through the Trotter product formula. Subsequently, the stochastic formulation in complex space is established, and applied to semi-classical expansions, nonlinear oscillations, quantum dynamics on analytic manifolds, dissipative systems (i.e., those which are open in that they interact with an environment), tunneling, field theory, and computer simulations. Thus, the covered field is vast, justifying the author's intention to formulate exactly the mathematical problem but only outline the proofs.

However, in practice the level of precision falls well short of that. From the first page it is already conspicuous that literature citations have been executed sloppily. A discrete-time stochastic process is also called a Markov chain, as if the Markov property (which never is clearly defined) would be trivial in discrete time. The normalization of a Master equation's transition rate is given just the wrong way round. Strikingly, `classical mechanics' is formulated in terms of a flow on a manifold---in fact defining a generic classical dynamical system, where mechanics proper should have involved canonical coordinates $p_i, q_i$ or more abstractly a symplectic structure. As a result, the introduction of Hamilton--Jacobi theory, a cornerstone for the author's semi-classical formalism, becomes unmotivated and \emph{ad hoc}. A decent canonical set-up is finally given some 100 pages later in Chapter~8, i.e., after its use in the crucial stochastic approach to the Feynman integral (Ch.~6). The latter has issues of its own, for it proceeds in terms of a `quantum Hamilton--Jacobi function,' reducing to the usual one from Chapter~1 for $\hbar\rightarrow0$. However, the whole discussion in Chapter~6 concerns the semi-classical limit only, so that it is unclear if the distinction between classical and quantum Hamilton--Jacobi functions in fact is substantial.

A look at the bibliography shows that the author has published the bulk of the material underlying this book in the form of single-author papers. Especially in such a situation, a fresh editorial look at the manuscript would have been invaluable. A text of improved clarity and consistency would make it much easier to gauge the true power of the author's methods. In any case, the nature of the subject does not merit a hasty publication.

In conclusion, the way in which this book has been produced does not justify its high price. Because of its wide scope, it may be of interest to those interested in stochastic formulations of quantum physics, and having sufficient background or determination not to be wrong-footed by the presentation. The book's stated goal of introducing a wider audience to these stochastic methods is only partially met.

\end{document}